\documentclass[twocolumn]{revtex4}
\usepackage{epsfig}
\usepackage{amssymb}
\usepackage{amsmath}
\usepackage{amsfonts}
\usepackage{graphicx}
\usepackage[dvips]{color}

\newcommand{\R}{\mathbb{R}}

\newcommand{\Z}{\mathbb{Z}}

\newcommand{\fa}{\mathfrak{a}}

\newcommand{\fg}{\mathfrak{g}}

\newcommand{\fs}{\mathfrak{s}}

\newcommand{\fz}{\mathfrak{z}}

\newcommand{\fK}{\mathfrak{K}}

\newcommand{\cP}{\mathcal{P}}

\newcommand{\cT}{\mathcal{T}}

\newcommand{\be}{\begin{equation}}
\newcommand{\ee}{\end{equation}}
\newcommand{\bea}{\begin{eqnarray}}
\newcommand{\eea}{\end{eqnarray}}

\newcommand{\ed}{\end{document}}

\newcommand{\rx}{\mbox{\large$\mathrm{x}$}}

\newcommand{\bi}{\begin{itemize}}
\newcommand{\ei}{\end{itemize}}

\newcommand{\bce}{\begin{center}}
\newcommand{\ece}{\end{center}}

\begin{document}
\title{Spectral Singularities of Complex Scattering Potentials and
Infinite\\ Reflection and Transmission Coefficients at real
Energies}
\author{Ali~Mostafazadeh}
\address{Department of Mathematics, Ko\c{c}
University, Sar{\i}yer 34450, Istanbul, Turkey\\
amostafazadeh@ku.edu.tr}

\begin{abstract}

Spectral singularities are spectral points that spoil the
completeness of the eigenfunctions of certain non-Hermitian
Hamiltonian operators. We identify spectral singularities of complex
scattering potentials with the real energies at which the reflection
and transmission coefficients tend to infinity, i.e., they
correspond to resonances having a zero width. We show that a wave
guide modeled using such a potential operates like a resonator at
the frequencies of spectral singularities. As a concrete example, we
explore the spectral singularities of an imaginary
$\cP\cT$-symmetric barrier potential and demonstrate the above
resonance phenomenon for a certain electromagnetic waveguide.
\medskip

\hspace{4.8cm}{Pacs numbers: 03.65.-w, 03.65.Nk, 11.30.Er, 42.25.Bs,
42.79.Gn}
\end{abstract}

\maketitle

\section{Introduction}

Complex $\cP\cT$-symmetric potentials \cite{bender-prl-1998} that
have a real spectrum are interesting, because one can restore the
Hermiticity of the corresponding Hamiltonian and uphold unitarity by
modifying the inner product of the Hilbert space
\cite{p2-p3,bender-prl-2002,jpa-2004b}. It is usually believed that
one can similarly treat every non-Hermitian Hamiltonian $H$ that has
a real and discrete spectrum. This is actually true provided that
$H$ has a complete set of eigenvectors \cite{p2-p3}. For the cases
that the spectrum is discrete the lack of completeness is associated
with the presence of exceptional points. These correspond to
situations where two or more eigenvalues together with their
eigenvectors coalesce. This phenomenon is known to have physically
observable consequences \cite{ep}. It also plays an important role
in the study of open quantum systems particularly in relation with
the resonance states \cite{nn,muller-rotter}. For the cases that the
spectrum has a continuous part, there is another mathematical
obstruction for the completeness of the eigenvectors called a
``spectral singularity'' \footnote{Unlike exceptional points,
spectral singularities are not associated with the coalescence of
eigenfunctions. Indeed, regardless of the presence of a spectral
singularity, to each point in the continuous spectrum there
corresponds two linearly independent eigenfunctions \cite{p88}.}.
The purpose of the present article is to describe the physical
meaning and a possible practical application of spectral
singularities.

Spectral singularities were discovered by Naimark \cite{naimark} and
subsequently studied by mathematicians in the 1950's and 1960's
\cite{ss-math}. The mechanism by which spectral singularities spoil
the completeness of the eigenfunctions and their difference with
exceptional points are discussed in \cite{p88}.

Spectral singularities of complex $\cP\cT$-symmetric and
non-$\cP\cT$-symmetric scattering potentials have been studied in
\cite{samsonov,p88}. In this article we shall examine the spectral
singularities of the imaginary potential \cite{rdm,jmp-2005}:
    \be
    v_{\fa,\fz}(\rx)=
    \left\{\begin{array}{cc}
    i\fz &{\rm for}~ -\fa<\rx<0,\\
    -i\fz &{\rm for}~0<\rx<\fa,\\
    0 & {\rm otherwise},\end{array}\right.
    \label{barrier-pot}
    \ee
with $\fa\in\R^+$ and $\fz\in\R-\{0\}$, that has applications in
modeling certain electromagnetic waveguides \cite{rdm}.

\section{Spectral singularities}

Consider a complex scattering potential $v(x)$ that decays rapidly
as $|x|\to\infty$ \footnote{Specifically, suppose that
$\int_{-\infty}^\infty (1+|x|)|v(x)|dx<\infty$.}. Suppose that the
continuous spectrum of the Hamiltonian $H=-\frac{d^2}{dx^2}+v(x)$ is
$[0,\infty)$, and for each $k\in\R^+$ let $\psi_{k\pm}(x)$ denote
the solutions of the eigenvalue equation $H\psi(x)=k^2\psi(x)$
satisfying the asymptotic boundary conditions:
    \be
    \psi_{k\pm}(x)\to e^{\pm ikx}~~{\rm as}~~x\to\pm\infty,
    \label{jost}
    \ee
i.e., the Jost solutions. A spectral singularity of $H$ (or $v$) is
a point $k_\star^2$ of the continuous spectrum of $H$ such that the
$\psi_{k_\star\pm}$ are linearly-dependent, i.e., they have a
vanishing Wronskian, $\psi_{k_\star+}\psi'_{k_\star-}-
\psi_{k_\star-}\psi'_{k_\star+}=0$, \cite{p88}.

Clearly the continuous spectrum of $H$ is doubly-degenerate. To make
this explicit, we use $\psi_{k}^{\fg}$ with $k\in\R^+$ and
$\fg\in\{1,2\}$ to denote a general solution of the eigenvalue
equation $H\psi(x)=k^2\psi(x)$. Furthermore, because $v(x)\to 0$ as
$x\to\pm\infty$, we have
    \be
    \psi_{k}^{\fg}\to A^\fg_\pm e^{ikx}+B^\fg_\pm e^{-ikx}
    ~~{\rm as}~~x\to\pm\infty,
    \label{asym}
    \ee
where $A^\fg_\pm$ and $B^\fg_\pm$ are complex coefficients. A
quantity of interest is the transfer
matrix $\mathbf{M}(k)$ that is defined by $\left(\begin{array}{c} A^\fg_+\\
B^\fg_+\end{array}\right)=\mathbf{M}(k)\left(\begin{array}{c} A^\fg_-\\
B^\fg_-\end{array}\right)$. Among its useful properties are the
identity $\det\mathbf{M}(k)=1$ and the following theorem.
    \begin{center}\parbox{8cm}{\textbf{Theorem~1:}
    \emph{$k_\star^2\in\R^+$ is a spectral singularity of $H$ if and only if
    either $-k_\star$ or $k_\star$ is a real zero of the entry $M_{22}(k)$ of
    $\mathbf{M}(k)$}, \cite{p88}.}\end{center}

Next, consider the left- and right-going scattering solutions of
$H\psi(x)=k^2\psi(x)$ that we denote by $\psi_k^l$ and $\psi_k^r$,
respectively. They satisfy \cite{muga}
    \bea
    \psi_k^l(x)&\to& \left\{
    \begin{array}{cc} N_l\left(e^{ikx}+R^l e^{-ikx}\right)&~{\rm
    as}~x\to-\infty,\\
    N_l T^l e^{ikx}&~{\rm
    as}~x\to+\infty,\end{array}\right.
    \label{psi-R}\\
    \psi_k^r(x)&\to& \left\{
    \begin{array}{cc}
    N_r T^r e^{-ikx}&~{\rm
    as}~x\to-\infty,\\
    N_r\left(e^{-ikx}+R^r e^{ikx}\right)&~{\rm
    as}~x\to+\infty,\end{array}\right.
    \label{psi-L}
    \eea
where $N_l,N_r,R^l,R^r,T^l$ and $T^r$ are complex coefficients.
$N_l,N_r$ are normalization constants, $|R^l|^2,|R^r|^2$ are the
left and right reflection coefficients, and $|T^l|^2,|T^r|^2$ are
the left and right transmission coefficients, respectively.
Comparing (\ref{psi-R}) and (\ref{psi-L}) with (\ref{jost}), we see
that $\psi_k^l$ and $\psi_k^r$ are respectively proportional to the
Jost solutions $\psi_{k+}$ and $\psi_{k-}$. Therefore, at a spectral
singularity, $k_\star^2$, the scattering solutions $\psi_k^l$ and
$\psi_k^r$ become linearly-dependent. In view of (\ref{psi-R}) and
(\ref{psi-L}), this is possible only if $R^l,R^r,T^l$ and $T^r$ tend
to infinity as $k\to k_\star$. The converse of this statement is
also true:
    \begin{center}\parbox{8.3cm}{\textbf{Theorem~2:}
    \emph{$k_\star^2\in\R^+$ is a spectral singularity of $H$ if and only if the
    left and right reflection and transmission coefficients tend to
    infinity as
    $k\to k_\star$ or $k\to -k_\star$.}}
    \end{center}
The following is an explicit proof of this theorem.

Comparing (\ref{psi-R}) and (\ref{psi-L}) with (\ref{asym}), we can
determine the coefficients $A^\fg_\pm$ and $B^\fg_\pm$ for
$\psi_k^l$ and $\psi_k^r$ and use them to express $R^l,R^r,T^l$ and
$T^r$ in terms of the entries of the transfer matrix
$\mathbf{M}(k)$. This yields
    \bea
    T^l&=&1/M_{22}(k),
    ~~~~R^l=- M_{21}(k)/M_{22}(k),
    \label{T-R-ell}\\
    T^r&=&1/M_{22}(k),
    ~~~~R^r= M_{12}(k)/M_{22}(k),
    \label{T-R-r}
    \eea
where we have employed $\det\mathbf{M}(k)=1$. As seen from
(\ref{T-R-ell}) and (\ref{T-R-r}), at a spectral singularity, where
$M_{22}(k)$ vanishes, $R^l,R^r,T^l$ and $T^r$ diverge. The converse
holds because $M_{12}$ and $M_{21}$ are entire functions (lacking
singularities).

Another curious consequences of (\ref{T-R-ell}) and (\ref{T-R-r}),
is the identity: $T^l=T^r$. This is derived in \cite{ahmed} using a
different approach, but is usually overlooked. See for example
\cite{muga}.

Next, we examine the $S$-matrix of the system:
    $\mathbf{S}=\left(\begin{array}{cc}
    T^l & R^r\\
    R^l & T^r\end{array}\right)$, \cite{muga}.
In view of (\ref{T-R-ell}), (\ref{T-R-r}), and
$\det\mathbf{M}(k)=1$, the eigenvalues of $\mathbf{S}$ are given by
$s_\pm=(1\pm\sqrt{1-M_{11}(k)M_{22}(k)})/M_{22}(k)$. At a spectral
singularity $s_+$ diverges while $s_-\to M_{11}(k)/2$. This suggests
identifying spectral singularities with certain type of resonances.
Indeed, in view of Theorem~2 and Siegert's definition of resonance
states \cite{siegert}, they correspond to resonances with a
vanishing width (real energy).

\section{$\cP\cT$-symmetric barrier potential}

Consider the Hamiltonian operator
$H=-\frac{d^2}{dx^2}+v_{\fa,\fz}(x)$ with $v_{\fa,\fz}(x)$ given by
(\ref{barrier-pot}). Because $v_{\fa,\fz}(x)=0$ for $|x|>\fa$, the
results of Section~II apply to $v_{\fa,\fz}$. The determination of
the eigenfunctions \cite{jmp-2005,ersan} of $H$ and the
corresponding transfer matrix $\mathbf{M}(k)$ is a straightforward
calculation. Here we report the result of the calculation of
$M_{22}(k)$:
    \be
    M_{22}(k)= e^{2i\fa k} [f_1(k)-if_2(k)]/\sqrt{1+y^2},
    \label{M22=}
    \ee
where $f_1$ and $f_2$ are real-valued functions given by
    \bea
    f_1(k)&=&\sqrt{1+y^2}\:\left|\cos(\fa kw)\right|^2
    -\left|\sin(\fa kw)\right|^2,
    \label{f1=}\\
    f_2(k)&=&\Re\left[\sqrt{1+iy}(2-iy)\sin(\fa kw)
    \cos(\fa kw^*)\right],~~~~~~~
    \label{f2=}
    \eea
$y:=\fz/k^2$, $w:=\sqrt{1-iy}$, and $\Re$ means ``real part of''.

According to Theorem~1 and Eq.~(\ref{M22=}), $k^2\in\R^+$ is a
spectral singularity of $v_{a,\fz}$ if and only if $f_1(k)=0$ and
$f_2(k)=0$. If we insert (\ref{f1=}) and (\ref{f2=}) in these
equations and divide their both sides by $|\cos(\fa k w)|^2$, we
find
    \bea
    \left|\tan(\fa kw)\right|^2&=&\sqrt{1+y^2},
    \label{f1=2}\\
    \tan(\fa kw)&=&
    -\left[\frac{\sqrt{1-iy}(2+iy)}{\sqrt{1+iy}(2-iy)}\right]\tan(\fa kw)^*.~~~~
    \label{f1=1}
    \eea
Now, we multiply both sides of (\ref{f1=1}) by $\tan(\fa kw)$, use
(\ref{f1=2}), $\cos(2\theta)=(1-\tan^2\theta)/(1+\tan^2\theta)$ and
$w=\sqrt{1-iy}$, to obtain
    $
    \cos(2\fa k\sqrt{1-iy})=-(1+4y^{-2})+2iy^{-1}.
    $
This equation is equivalent to
    \bea
    \cos r\:\cosh q&=&-(1+4y^{-2}),
    \label{eq1}\\
    \sin r\:\sinh q&=&2y^{-1},
    \label{eq2}
    \eea
where
    \bea
    q&:=&
    \fa k~\sqrt{2\left(\sqrt{y^2+1}-1\right)}~{\rm sgn}(y),
    \label{q-def}\\
    r&:=&
    \fa k~\sqrt{2\left(\sqrt{y^2+1}+1\right)},
    \label{r-def}
    \eea
${\rm sgn}(y)$ denotes the sign of $y$, and we have employed the
identities $\sin(\mbox{$\frac{\tan^{-1}y}{2}$})={\rm sgn}(y)
    \sqrt{\frac{1}{2}\left[1-(y^2+1)^{-1/2}\right]}$ and
$\cos(\mbox{$\frac{\tan^{-1}y}{2}$})=
    \sqrt{\frac{1}{2}\left[1+(y^2+1)^{-1/2}\right]}$.

Next, we solve for $y^{-1}$ in (\ref{eq2}), substitute the resulting
expression in (\ref{eq1}), and use the identities $\sinh^2
q=\cosh^2q-1$ and $\cos^2r=1-\sin^2r$ to obtain a quadratic equation
for $\cosh q$ with solutions
    \be
    \cosh q=\frac{1}{2}(-1\pm\sqrt{2\cos(2r)-1})\cot r~\csc r.
    \label{cos-q}
    \ee
To ensure that the right-hand side of this equation is real, we must
have $\cos(2r)\geq \frac{1}{2}$. Furthermore according to
(\ref{eq1}), $\cos(r)< 0$. These imply
    \be
    |r-(2n+1)\pi|\leq \frac{\pi}{6},~~~\mbox{for some integer $n$.}
    \label{condi}
    \ee
Under this condition the right-hand side of (\ref{cos-q}) is greater
than 1. Hence, $q=\pm q_\pm(r)$, where
    $ q_\pm(r):=\cosh^{-1}\left[ \cot r\,\csc r
    (\pm\sqrt{2\cos(2r)-1}-1)/2\right].$
If we set $q=\pm q_\pm(r)$ in (\ref{eq2}) and solve for $y$, we find
$y=\pm\,{\rm sgn}(\sin r)y_\pm(r)$, where $y_\pm:=2|\sin r\,\sinh
q_\pm (r)|^{-1}$. Inserting this expression for $y$ in (\ref{q-def})
and (\ref{r-def}) and solving for $q$ give $q=\pm \tilde
q_{\pm}(r)$, where $\tilde q_\pm(r):=r\:{\rm sgn}(\sin r) \:
\sqrt{\frac{\sqrt{y_\pm(r)^2+1}-1}{
    \sqrt{y_\pm(r)^2+1}+1}}.$
The spectral singularities correspond to the values of $r$ for which
$q_\pm(r)=\tilde q_\pm(r)$. These are transcendental equations
admitting simple numerical treatments. It turns out that
$q_+(r)=\tilde q_+(r)$ does not have a real solution fulfilling
(\ref{condi}), while $q_-(r)=\tilde q_-(r)$ has two solutions $\pm
r_n$ for each choice of $n$ in (\ref{condi}). Table~1 lists the
numerical values of $r_n$ for various choices of $n$. It turns out
that $r_n>0$ and $r_{-n}=-r_{n+1}$ for $n>0$.
  \begin{table}
  {\small\begin{tabular}{|c|c|c|c|c|}
  \hline
  $n$ & $r_n$ & $y_{n}$ & $\fa k_n$ & $\fa^2\fz_n$ \\
  \hline \hline
  0& 2.64390700 & 1.82765566 & 1.06468255 & 2.07173713\\
  \hline
  1& 9.11655393 & 0.71364271 & 4.31823693 & 13.3074170\\
  \hline
  2& 15.4804556 & 0.49008727 & 7.52928304 & 27.7830976\\
  \hline
  10& 65.8884385 & 0.17167639 & 32.8243878 & 184.971084\\
  \hline
  100& 631.445619 & 0.02901727 & 315.689592 & 2891.85852\\
  \hline
    \end{tabular}}
    \caption{$r_n$, $y_{n}$, $k_n$, and $\fz_n$ are respectively the
    numerical values of $r,y_-,k$ and $\fz$ that correspond to
    spectral singularities. For $n\geq 0$, $k_n$ and $\fz_n$ are
    increasing functions of $n$.
    \label{table1}}
    \end{table}

Next, we insert $y=\pm\,{\rm sgn}(\sin r)y_-(r)$ in (\ref{r-def})
and use the identity $\fa^2\fz=(\fa k)^2 y$ to obtain $k=g(r)$ and
$\fa^2\fz=\pm g(r)^2\,{\rm sgn}(\sin r)\,y_-(r)$, where $g(r):=
r/\sqrt{2 (\sqrt{y_-(r)^2+1}+1 )}$. Setting $r=\pm r_n$ in these
relations gives the values ($\fa k_n$ and $\fa^2\fz_n$) of $\fa k$
and $\fa^2 \fz$ that are associated with spectral singularities. We
list some of these values in Table~\ref{table1}. Because $\fa k$ and
$\fa^2 \fz$ are odd functions of $r$ and $r_{-n}=-r_{n+1}$ for
$n>0$, we have $k_{-n}=-k_{n+1}$ and $\fz_{-n}=-\fz_{n+1}$.
According to Table~\ref{table1}, the smallest values of $\fa |k|$
and $\fa^2|\fz|$ for which a spectral singularity occurs are
respectively $\fa k_0\approx 1.06$ and $\fa^2 \fz_0\approx 2.07$.
Using more accurate values for $\fa k_n$ and $\fa^2\fz_n$ that we do
not report here, we have checked that $|M_{22}(k_n)|<10^{-9}$ for
$|n|\leq 20$ and $|n|=10^2,10^3,10^4$.

For the system we considered in this section, each value of
$\fa^2\fz$ can support at most one spectral singularity (either the
latter does not exist or it exists for a single energy value).

\section{A $\cP\cT$-symmetric waveguide}

\begin{figure}
\begin{center} 
\includegraphics[scale=.80,clip]{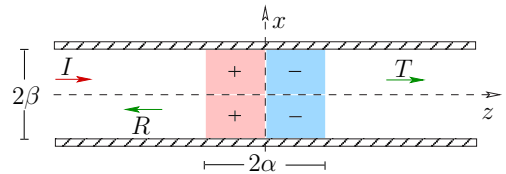}\vspace{-.3cm}
\caption{Cross section of a waveguide with gain (+) and loss (-)
regions in the $x$-$z$ plane. Arrows labeled by $I$, $R$, and $T$
represent the incident, reflected, and transmitted
waves.\label{fig1s}}
\end{center}
\end{figure}

Consider a rectangular waveguide with perfectly conducting walls
that is aligned along the $z$-axis and has height $2\beta$ as
depicted in Figure~\ref{fig1s}. Suppose that the region $|z|<\alpha$
inside the waveguide is filled with an atomic gas, and a laser beam
shining along the $y$-direction in the region $-\alpha<z<0$ is used
to excite the resonant atoms and produce a population inversion. In
this way $-\alpha<z<0$ and $0<z<\alpha$ serve as gain and loss
regions respectively, and the relative permittivity at the resonance
frequency takes the form $\varepsilon(z)=1+\frac{i\omega^2_p~{\rm
sgn}(z)}{2\delta\omega}$ for $|z|<\alpha$ and $\varepsilon(z)=1$ for
$|z|\geq\alpha$, where $\omega,\omega_p$ and $\delta$ are
respectively the frequency of the wave, plasma frequency, and the
damping constant, \cite{rdm}. Alternatively,
$\varepsilon(z)=1-v_{\alpha,\fs/\omega}(z)$ where
$\fs:=\omega^2_p/(2\delta)$. In \cite{rdm}, the authors use an
approximation scheme to reduce Maxwell's equations for this system
to the Schr\"odinger equation for the barrier
potential~(\ref{barrier-pot}). Here we offer an exact treatment to
examine singularities of the reflection and transmission
coefficients for this waveguide.

Let $\hat i,\hat j,\hat k$ be the unit vectors along the $x$-, $y$-
and $z$-axes, $\fK:=\omega/c$, $m\in\Z^+$, $\fK_m:=\pi m/(2\beta)$,
$\chi_m(x)):=\sin[\fK_m(x+\beta)]$, and
$\kappa:=\sqrt{\fK^2-\fK_m^2}$. Then
    $
    \vec E(\vec r,t)=\Re \left(e^{-i\omega t}[-i\omega
    \chi_m(x)\phi(z)]\hat j\right),
    $ and
    $\vec B(\vec r,t)=\Re \left(e^{-i\omega t}[
    \chi_m(x)\phi'(z)\hat i-\chi_m'(x)\phi(z)\hat k]\right),~~~~
    $
are transverse electric (TE) waves satisfying the boundary
conditions for the waveguide and solving Maxwell's equations
provided that
    \be
    \phi''(z)+[\fK^2\varepsilon(z)-\fK_m^2]\phi(z)=0,
    \label{eq}
    \ee
and $\phi$ and $\phi'$ are continuous functions on the $z$-axis
\footnote{This is true independently of the width of the
waveguide.}. For $|\fK|>\fK_n$, the solution of (\ref{eq}) has the
form (\ref{asym}) with $z$ and $\kappa$ playing the roles of $x$ and
$k$ respectively. This allows us to define a transfer matrix
$\mathbf{M}(\kappa)$ for this system and introduce the right and
left transmission and reflection amplitudes, $T^{l,r}$ and
$R^{l,r}$, associated with (\ref{eq}). These satisfy (\ref{T-R-ell})
and (\ref{T-R-r}), and diverge whenever $M_{22}(\kappa)=0$.

It is not difficult to see that the right and left reflection and
transmission amplitudes for the propagating TE wave coincide with
$T^{l,r}$ and $R^{l,r}$, respectively. Therefore, if we can tune the
frequency $\omega$ of the incoming wave to the frequency
$\omega_\star$ of a spectral singularity, then the amplitude of the
wave will diverge as $\omega\to\omega_\star$. In practice, this
means that sending in a wave of frequency
$\omega\approx\omega_\star$ will induce outgoing (transmitted and
reflected) waves of considerably enhanced amplitude. The waveguide
then uses a part of the energy of the laser beam to produce and emit
a more intensive electromagnetic wave. Note that this effect is
fundamentally different from the resonance effects associated with
exciting resonance modes of a cavity resonator. Unlike the latter
that has a geometric origin, the spectral singularity-related
resonance effect relies on the existence of a localized region with
a complex permittivity (a complex scattering potential).

\begin{figure}
\begin{center} 
\includegraphics[scale=.35,clip]{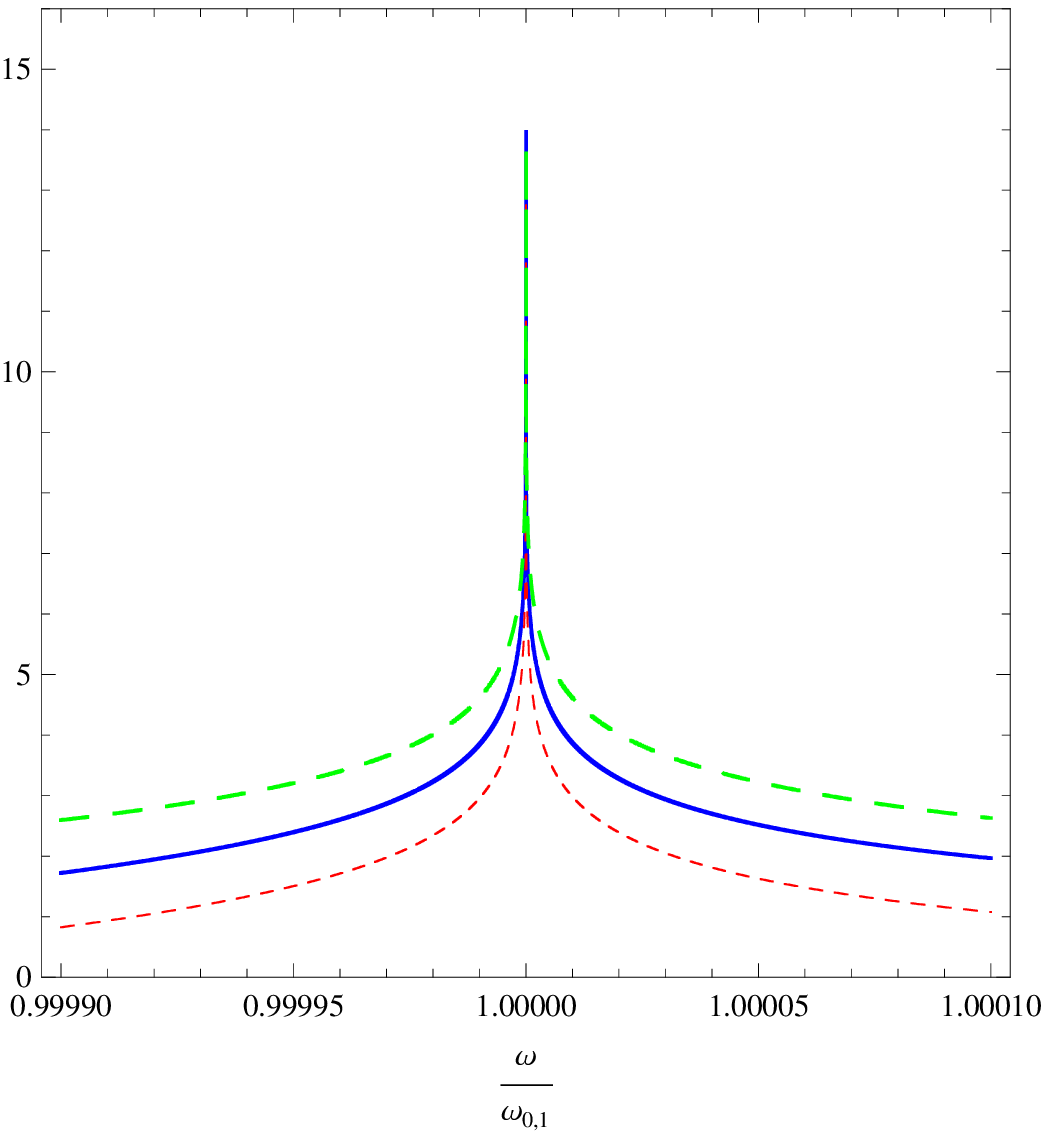}~~
\includegraphics[scale=.35,clip]{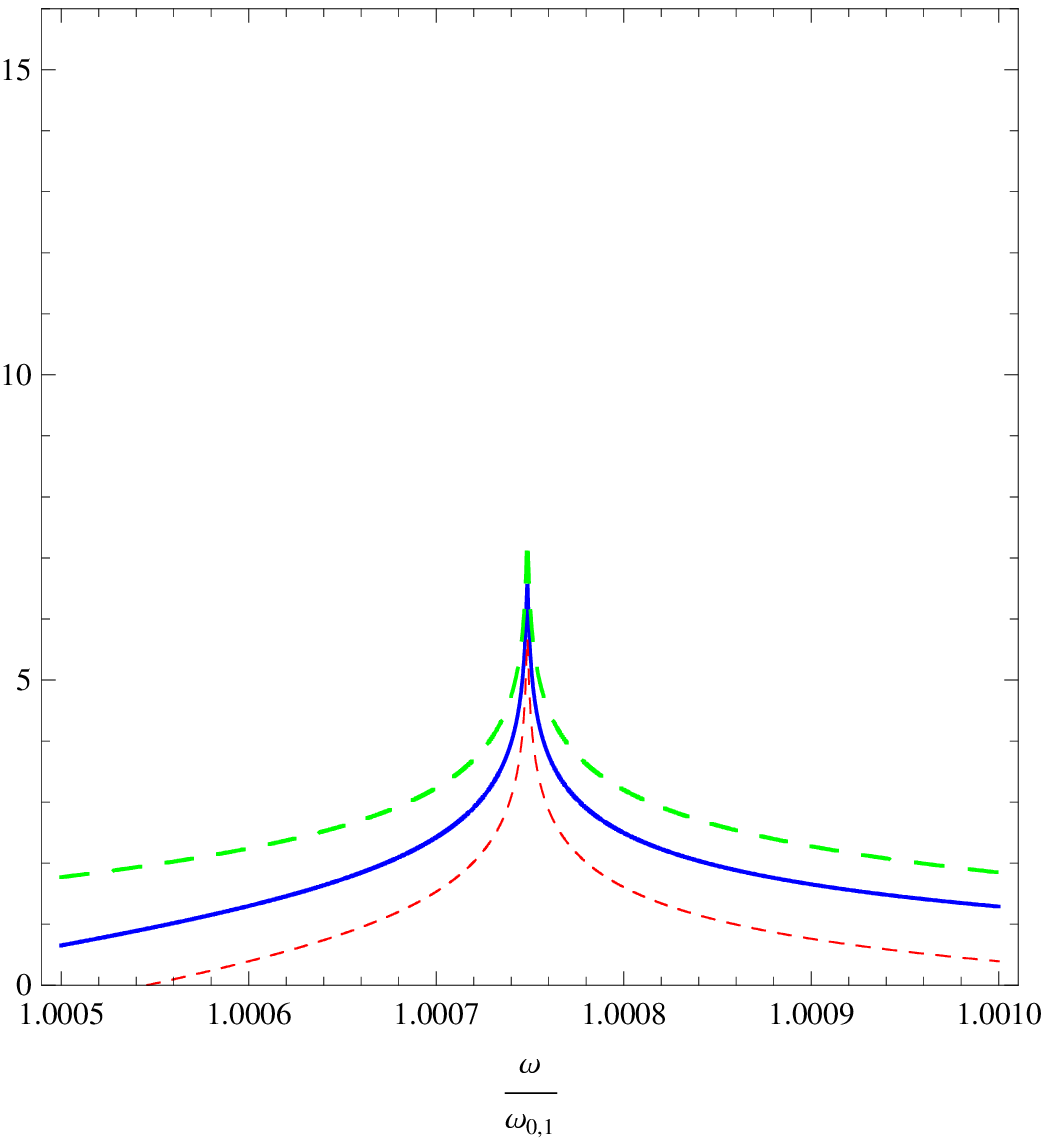}\vspace{-.3cm}
{\caption{Graphs of $\log_{10}(|T^{r,l}|^2)$ (full blue curves),
$\log_{10}(|R^{l}|^2)$ (dotted red curves), and
$\log_{10}(|R^{r}|^2)$ (dashed green curves) as a function of
$\omega/\omega_{0,1}$, for $m=1$, $n=0$, $\hbar\omega_{0,1}=5~{\rm
eV}$, $\hbar\omega_p=0.2~{\rm eV}$, $\hbar\delta=1.25~{\rm eV}$. For
the figure on the left (right) $\alpha= 1004.17~{\rm n m}$, $\beta=
62.0464~{\rm nm}$ ($\alpha= 1004~{\rm n m}$, $\beta= 62~{\rm
nm}$).\label{fig1}}}
\end{center}
\end{figure}

The calculation of the transfer matrix $\mathbf{M}(\kappa)$ defined
by (\ref{eq}) is analogous to that of the $\cP\cT$-symmetric barrier
potential. In fact, $M_{22}(\kappa)$ takes the form (\ref{M22=})
provided that we set: $k=\kappa$, $\fa=\alpha$, and
$y=\fs\fK/(c\kappa^2)$. In particular, we can determine the values
of $\omega$ and $\fs$ for which the above resonance phenomenon
occurs by setting $\kappa=k_n$ and $\fs\fK/(c\kappa^2)=\pm y_{n}$.
This yields $\omega=\omega_{n,m}$ and $ \fs=\fs_{n,m}$ where $
\omega_{n,m}:=c\sqrt{k_n^2+\fK_m^2}$ and $\fs_{n,m}:= \pm c\,
k_n^2y_{n}/\sqrt{k_n^2+\fK_m^2}= \pm c^2\fz_n/\omega_{n,m}.$ For
$m=1$, $\hbar\omega_p=0.2~{\rm eV}$, $\hbar\delta=1.25~{\rm eV}$, we
attain the spectral singularity with $n=0$ for
$\hbar\omega=\hbar\omega_{0,1}=5~{\rm eV}$, $\alpha\approx 1004~{\rm
n m}$ and $\beta\approx 62~{\rm nm}$. Figure~\ref{fig1} shows the
graphs of the logarithm of the transmission and reflection
coefficients as functions of $\omega/\omega_{0,1}$ for this case.
The location and hight of the pick representing the spectral
singularity are highly sensitive to the values of $\alpha$ and
$\beta$.

We close this section by noting that similar $\cP\cT$-symmetric
waveguides have been considered in \cite{kgm,periodic}. These differ
from the one we studied in that in our case the permitivity changes
along the direction of the propagation of the wave ($z$-axis). This
is essential for realizing the spectral singularity-related
resonance effect.

\section{Concluding Remarks}

In this article, we offered for the first time a simple physical
interpretation for the spectral singularities of complex scattering
potentials. In the framework of pseudo-Hermitian quantum mechanics
\cite{jpa-2004b}, where one defines unitary quantum systems with a
non-Hermitian Hamiltonian by modifying the inner product of the
Hilbert space, the presence of spectral singularities is an
unsurmountable obstacle \cite{p88}. In contrast, in the standard
applications of non-Hermitian Hamiltonians, spectral singularities
are interesting objects to study, because they correspond to
scattering states (with real energy) that nevertheless behave like
resonant states.

We explored the spectral singularities of a $\cP\cT$-symmetric
potential $v_{\fa,\fz}$ that admits a realization in the form of a
waveguide. We obtained the values of the physical parameters of the
waveguide and the propagating TE wave for which the system displays
the resonance behavior associated with the spectral singularities of
$v_{\fa,\fz}$.

Our results call for a more extensive investigation of the spectral
singularities of complex scattering potentials that can be realized
experimentally. This should provide means for the observation of the
resonance effect that is foreseen in this article. Another line of
research is to explore the spectral singularities of complex
periodic potentials \cite{gasymov}. A more basic problem is to study
the consequences of spectral singularities for the implementation of
quantum scattering theory. Similarly to exceptional points, presence
of spectral singularities leads to subtleties associated with the
existence of an appropriate resolution of identity \cite{sokolov}.
This problem may be avoided for scattering wave packets obtained by
superposing eigenfunctions of the Hamiltonian. A general treatment
of this problem requires a separate investigation.

\textbf{Acknowledgements:} I wish to thank Hossein Merhi-Dehnavi for
helpful discussions. This work was supported by the Scientific and
Technological Research Council of Turkey (T\"UB\.{I}TAK) in the
framework of the project no: 108T009 and by Turkish Academy of
Sciences (T\"UBA).

\ed